\documentclass[prb,showpacs,preprintnumbers,amsmath,amssymb]{revtex4}

\usepackage{graphicx,color,psfrag}
\usepackage{dcolumn}
\usepackage{bm}

\begin{document}


\title{Exciton Mediated One Phonon Resonant Raman Scattering from One-Dimensional Systems}

\author{A. N. Vamivakas$^1$}
\email{nvami@bu.edu}
\author{A. Walsh$^2$}%
\author{Y. Yin$^2$}
\author{M. S. \"{U}nl\"{u}$^1$}
\author{B. B. Goldberg$^2$}
\author{A. K. Swan$^1$}
\affiliation{%
$^1$Department of Electrical and Computer Engineering, Boston
  University, Boston, Massachusetts 02215, USA\\
$^2$Department of Physics, Boston
  University, Boston, Massachusetts 02215}%


\date{\today}
\begin{abstract}
 We use the Kramers-Heisenberg approach to derive a general expression for
 the resonant Raman scattering cross section from a one-dimensional (1D) system
 explicitly  accounting for excitonic effects. The result should prove useful for analyzing the
 Raman resonance excitation profile lineshapes for a variety of 1D
 systems including carbon nanotubes and semiconductor quantum wires.  
We apply this formalism to a simple 1D model system to illustrate
 the similarities and differences between the free electron and
 correlated  electron-hole theories.
\end{abstract}


\maketitle

\section{\label{sec:level1}Introduction}

  Raman scattering is a standard optical spectroscopy technique used to
  characterize the excitation spectrum of a material system.  If the
  exciting or scattered light frequency is nearly commensurate with an
  electronic transition of the material, the scattered Raman signal
  intensity is greatly enhanced \cite{CardScatt}.  In semiconductors or
  insulators, resonant Raman scattering not only serves as a probe of
  a structure's vibrational modes, but also can provide valuable
  information about the nature of a material's electronic structure.

  For three-dimensional (3D) bulk semiconductors \cite{Card89a},
  two-dimensional (2D)
  quantum wells \cite{Tang89a,Ploo96a}, zero-dimensional (0D) self-assembled
  quantum dots \cite{Ullo99a} and semiconductor microcrystallites
  \cite{Card95a}, a Kramers-Heisenberg approach to the theory of one
  phonon resonant Raman scattering (1phRRS), based on either free
  electron-hole states (FEH) or Wannier excitonic states have been
  well developed.  But, to our knowledge, a theory of 1phRRS
  incorporating excitonic effects has not been developed for a
  quantum confined one-dimensional (1D) system.  Here we construct an expression for the
  resonant Raman scattering cross-section from a 1D system
  incorporating Wannier excitons as the intermediate electronic states.  Specifically, we
  derive a general expression that is useful for analyzing the
  Raman spectra of a variety of 1D systems.

  There are currently two approaches to achieving quantum confinement
  that results in a 1D system.  One approach, in the spirit
  of quantum wells and quantum dots, involves introducing a potential
  discontinuity to confine the material's electronic quasiparticles along the direction of discontinuity.  A second
  approach, exemplified by single wall carbon nanotubes (SWNTs), is
  geometric confinement where a periodic boundary condition is used to
  provide single-valued solutions.  For example, by rolling a graphene sheet to form a continuous tube, the
  allowable states of the system must exhibit phase continuity as we
  traverse a complete cycle around the tube in the azimuthal
  direction. 

 SWNTs are of particular interest since there has been decisive theoretical
 \cite{Ando97a,Avou04a,Mele04a} and experimental
 \cite{Hein05a,Lien05a} evidence that excitonic states dominate the
 optical properties of these systems.  In studying SWNT, 1phRRS is a
 standard optical technique utilized to locate electronic
 resonances in SWNTs and to determine the diameter of the tube
 under study.  With the ability to perform tunable 1phRRS on a single
 SWNT and map out
 the full Raman scattering resonant excitation profile (REP)
 \cite{Swan05a}, a theory of 1phRRS that explicitly accounts for the
 excitonic intermediate states of the scattering process is necessary.

 The organization of this paper is as follows.  In Section II, we
 develop a general expression for the 1phRRS cross-section of a
 1D system. First, we solve a 1D Schr\"{o}dinger
 equation for the Wannier exciton wavefunctions and energy
 eigenvalues.  We use the wavefunctions to construct the appropriate interaction
 Hamiltonians necessary to describe the Raman scattering process.  In Section III, for the particular case of a
 two-subband model, we illustrate the influence of
 excitonic states on the 1phRRS cross-section. Finally, in
 Section IV we present a summary of the work and a comparison between
 1phRRS and 1D absorption.  

 \section{Theory of One Phonon Raman Scattering from a One-Dimensional System}

\begin{figure}[t]
\begin{center}
\includegraphics[width=3.34in]{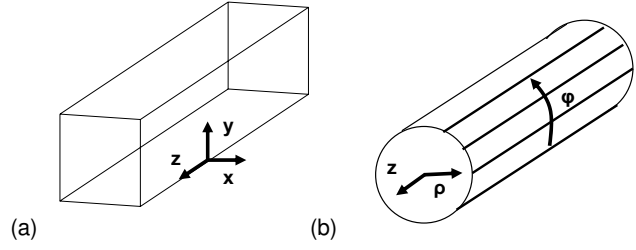}
\end{center}
\caption{Two typical one-dimensional quantum confined systems. (a) illustrates a
  rectangular quantum wire and (b) a cylindrical tube.  }
\label{Fig1}
\end{figure}

The general 1D material system we consider is illustrated in
Fig. \ref{Fig1}.
We imagine the system is illuminated by a laser beam of fixed
frequency ${\omega_l}$, propagation direction $\vec{q}_l$ and polarization $\vec{e}_l$.  The
inelastically 
 scattered radiation propagates in direction $\vec{q}_s$
with fixed polarization $\vec{e}_s$ and is collected so that its spectral
content, $\omega_s$, may be analyzed with a spectrometer.  Without loss of
generality, we focus on the Stokes scattering process.
Microscopically,  an incident pump photon interacts with the unexcited
system,
creates an electronic excitation that scatters a  phonon before
relaxing radiatively back to its ground state by emitting a photon.
We use Fermi's golden rule to determine the Stokes
differential Raman scattering cross-section integrated over all
scattered photon wavenumbers. 
The resulting
 expression is \cite{CardScatt} 

\begin{equation}
\label{RRSCS}
\frac{d\sigma_{RRS}}{d\Omega} =
\frac{\omega_s^3n_s^3n_lV^2_{crystal}}{\omega_lc^4(2\pi\hbar)^2}\cdot
|W_{i\rightarrow f}
(\omega_l,\vec{e_l};\omega_s=\omega_l-\Omega_p,\vec{e_s})|^2
\end{equation}

\noindent where $c$ is the speed of light in free space, $n_i$ is the
refractive index of the material evaluated at frequency $\omega_i$, $V_{crystal}$ is the volume of the material system
and $|W_{i \rightarrow f}|^2$ is the transition probability from
initial system state $i$, with a single pump photon, to a final state
$f$, with a single scattered photon and a single phonon.
Using third order time-dependent perturbation theory,
the transition matrix element $|W_{i \rightarrow f}|^2$ from initial state $|initial>$ to final
state $|final>$ can be expressed as

\begin{equation}
\label{MatEl}
W_{i \rightarrow f} = \sum_{a,b}\frac{<final|\hat{H}_{int}|b><b|
  \hat{H}_{int} |a><a|\hat{H}_{int} |initial>}
  {(E_{initial}-E_{b}-i\Gamma_b)(E_{initial}-E_{a}-i\Gamma_a)}
\end{equation}

\noindent where the sum over $a$ and $b$ is over all permissible intermediate states,
$E_j$ 
is the energy associated with state $j$,
$\hat{H}_{int}$ is the appropriate interaction Hamiltonian (to be defined
below), $\Gamma_j$ is a phenomenological broadening
parameter that incorporates the finite lifetime of intermediate
states $j=a,b$ and $i$ is the imaginary unit $\sqrt{-1}$.  To evaluate
Eq. \eqref{MatEl}, we will use the language of
second quantization and specify the states in the occupation number
representation.  In this notation, a general state is a direct product
of state vectors where each component state vector belongs to the sector of the  Hilbert space
appropriate for the excitation;
$|state>=|electronic>\otimes|phonon>\otimes|photon>$. 
 We label each state by the
number of quanta in a given mode. For example,
$|photon>=|0_{\vec{q}_s,\vec{e}_s},1_{\vec{q}_l,\vec{e}_l}>$
when there is a single photon in the laser mode and there is no
scattered photon.
  The role of the interaction
Hamiltonian, $\hat{H}_{int}$, in this perturbative treatment is to allow quanta to be
exchanged between the different sectors of Hilbert space.
$\hat{H}_{int}$ can be expressed as
$\hat{H}_{int}=\hat{H}^{(X-R)}_{int}+\hat{H}^{(X-L)}_{int}$ where we have
decomposed the interaction Hamiltonian into a piece $\hat{H}^{(X-R)}_{int}$ that couples the
excitons with the photons (R=radiation) and a piece
$\hat{H}^{(X-L)}_{int}$ that couples the excitons with
the phonons (L=lattice).  
 Here it is important to note
 that the intermediate electronic excitations
will be treated as correlated electron-holes, or excitons, and not as
free electrons and holes.         

Upon substitution of $\hat{H}_{int}$ into Eq. \eqref{MatEl} and requiring
energy conservation, we find there
are six possible pathways or probability amplitudes that can
contribute to $|W_{i \rightarrow f}|^2$.  In the following we focus
only on the contribution of the  resonant path because the other five
pathways make a comparatively negligble contribution
to the cross-section. The contribution of the resonant pathway can be expressed as 

\begin{equation}
\label{Res}
W_{i \rightarrow f} = \sum_{a,b}\frac{<final|\hat{H}_{int}^{(X-R)}|b><b|
  \hat{H}_{int}^{(X-L)}|a><a|\hat{H}_{int}^{(X-R)}|initial>}
  {(\hbar\omega_l-\hbar\Omega_p-E_b-i\Gamma_b)(\hbar\omega_l-E_{a}-i\Gamma_a)}.
\end{equation}

\noindent To proceed further we define the exciton-radiation
interaction Hamiltonian $\hat{H}^{(X-R)}_{int}$, assuming the minimal coupling interaction and retaining
only the term linear in the electromagnetic vector potential, written in
second quantized notation as \cite{Card89a}

\begin{equation}
\label{XRint}
\hat{H}_{int}^{(X-R)}=\sum_{\substack{X,\vec{K}_{cm} \\\\ \vec{q},\vec{e}} } 
   T_{cv}^X(\vec{K}_{cm}) \hat{D}^{\dagger}_{X,\vec{K}_{cm}}
   \hat{a}_{\vec{q},\vec{e}} + T_{cv}^{*X}(\vec{K}_{cm})
   \hat{D}_{X,\vec{K}_{cm}} \hat{a}^{\dagger}_{\vec{q},\vec{e}}
\end{equation}

\noindent where the exciton-radiation coupling constant can be expressed as

\begin{equation}
\label{XR1}
T_{cv}^{X}(\vec{K}_{cm}=\vec{k}_e-\vec{k}_h)=\frac{e}{m_o}\sqrt{\frac{2\pi\hbar}{V_{crystal}\omega n^2
}}<X_2|e^{i\vec{q}\cdot\vec{r}}\vec{e}\cdot\vec{\hat{p}}|X_1>
\end{equation}

\noindent and 
$\hat{D}_{X,K_{cm}},\hat{D}^{\dagger}_{X,K_{cm}}$ ($(\hat{a}_{\vec{q},\vec{e}},\hat{a}^{\dagger}_{\vec{q},\vec{e}})$)
are the annhilation and creation operators for excitons (photons),
$e$ is the electronic charge, $m_o$ is the bare electron mass,
$\vec{\hat{p}}$ is the electronic momentum operator and $|X_j>$
are exciton wavefunctions.

Similarly, focusing only on phonon creation processes, the
exciton-lattice interaction Hamiltonian for coupling with a single
phonon branch can be expressed as

\begin{equation}
\label{XLint}
\hat{H}_{int}^{(X-L)}=\sum_{\substack{X_1,\vec{K}_{cm,1} \\\\ X_2,\vec{K}_{cm,2} \\\\ \vec{Q}}}
   S_{X_1}^{X_2}(\vec{Q}) \hat{D}^{\dagger}_{X_2,\vec{K}_{cm,2}} \hat{D}_{X_1,\vec{K}_{cm,1}}
   \hat{b}^{\dagger}_{\vec{Q}}
\end{equation}

\noindent where the exciton-lattice coupling constant is

\begin{equation}
\label{XL1}
S_{X_1}^{X_2}(\vec{Q}) =
<X_2|C(\vec{r}_{e})e^{-i\vec{Q}\cdot\vec{r}_e}-C(\vec{r}_{h})e^{-i\vec{Q}\cdot
  \vec{r}_h}|X_1>
\end{equation}

\noindent and $\vec{Q}$ is the phonon
momentum, $\hat{b}^{\dagger}_{\vec{Q}}$ is the phonon creation
operator, and  $C(\vec{r_j})$ depend on the details of the exact
exciton-phonon interaction.  Two common examples of exciton-phonon interactions are deformation
potential coupling and Fr{\"o}lich coupling \cite{PhonNano}.  To determine the explicit
form of both $\hat{H}_{int}^{(X-R)}$ and $\hat{H}_{int}^{(X-L)}$ we need expressions
for the 1D exciton wavefunctions $|X_j>$ so that we can
evaluate the appropriate coupling constants.

To obtain expressions for the exciton wavefunctions, we will use both the effective mass
approximation (EMA) and the envelope function approximation (EFA)\cite{OpPropSemi}.
These approximations reduce the complicated problem of solving the
Schr\"{o}dinger equation, for the two-particle Bloch wavefunction to
solving the following modified Schr\"{o}dinger equation \cite{Taka91a}

\begin{equation}
\label{EMAFull}
\Biggl[-\frac{\hbar^2}{2m_e^*}\nabla^2_e
  -\frac{\hbar^2}{2m_h^*}\nabla^2_h+V(\vec{r}_e)+V(\vec{r}_h)+V_3(\vec{r}_e,\vec{r}_h)\Biggr]
\Phi(\vec{r}_e,\vec{r}_h) = E \Phi(\vec{r}_e,\vec{r}_h)
\end{equation}

\noindent for the \textit{envelope function}
 $\Phi(\vec{r}_e,\vec{r}_h)$
 of the exciton where the influence of
the periodic crystal potential has been incorporated into the problem
by replacing the bare electron (hole) mass with the \textit{effective
  electron (hole)
  mass} $m_e^*$ ($m_h^*$) and by multiplying the solution of Eq. \eqref{EMAFull} by the Bloch functions of the conduction $u_c(\vec{r})$ and valence
band $u_v(\vec{r})$ (for the unconfined system) to calculate the approximate
two-particle Bloch wavefunction $\Psi(\vec{r}_e,\vec{r}_h) \approxeq \Phi(\vec{r}_e,\vec{r}_h)u_c(\vec{r}_e)u_v(\vec{r}_h)$.
In Eq. \eqref{EMAFull}, $V(\vec{r}_j)$ is the potential that confines
particle $j$ and $V_3(\vec{r}_e,\vec{r}_h)$ is the 3D
Coulomb potential.  Assuming we have solved the problem of electron and hole confinement, we can expand the exciton envelope
function in the basis of the confined electron and hole wavefunctions as
\begin{equation}
\label{ExEnv}
\Phi(\vec{r}_e,\vec{r}_h)=\sum_{\substack{(l_e^a,l_e^b,l_h^a,l_h^b)\\integers}} e^{iK_{cm}Z_{cm}}
  \phi_x^{(l_e^a,l_e^b,l_h^a,l_h^b)}(z_r=z_e-z_h)
f^{l_e^a}(x_e^a)f^{l_e^b}(x_e^b)f^{l_h^a}(x_h^a)f^{l_h^b}(x_h^b)
\end{equation}

\noindent where $x^j_i$ denotes the $j^{th}$ coordinate for particle
$i$, $f^{l_i^j}$ is the complete and orthonormal
confined wavefunction for particle $i$ in subband labeled by $l_i^j$
confined in the $j^{th}$ coordinate direction and, anticipating a change to center of mass coordinates along the
unconfined direction,
$\phi_x^{(l_e^a,l_e^b,l_h^a,l_h^b)}(z_r=z_e-z_h)$ is a function
characterizing the relative motion of the electron and hole.  
 In Eq. \eqref{ExEnv}, we have assumed that the confining potential is along two
spatial directions and unconfined motion is along a third orthogonal
direction.  For example, in Fig. \ref{Fig1}(a), electronic motion is confined
along the $x$ and $y$ directions while it is unconfined along the $z$
direction.  If we began with a material system that was initially
2D, and then further confined along one spatial direction,
 we would supress all functions and coordinates in Eq. \eqref{ExEnv}
 labeled with $b$.   
 Equation  \eqref{ExEnv}
expresses the two-particle envelope function as a superposition of confined
electron and hole states, weighted by a function describing
the 1D excitonic state associated with the
subbands $(l_e^a,l_e^b,l_h^a,l_h^b)$.  Finally, we substitute
Eq. \eqref{ExEnv} into Eq. \eqref{EMAFull} and project the resulting
equation over a set of confined wavefunctions with labels
$(l_e^a,l_e^b,l_h^a,l_h^b)$.  Here we neglect the
  possibility of subband coupling due to the Coulomb potential, a
  reasonable assumption for large subband energy spacing, and 
arrive at the following, effective 1D Schr\"{o}dinger
equation for the relative motion of the electron and hole

\begin{equation}
\label{Schro}
\Biggl[-\frac{\hbar^2}{2\mu}\frac{d^2}{d^2z_r}+V_{1-eff}^{(l_e^a,l_e^b,l_h^a,l_h^b)}(z_r)\Biggr]\phi_x^{(l_e^a,l_e^b,l_h^a,l_h^b)}(z_r)
= E_x^{(l^e_a,l^e_b,l^h_a,l^h_b)}(K_{cm})\phi_x^{(l_e^a,l_e^b,l_h^a,l_h^b)}(z_r)
\end{equation}



\noindent where  $\mu^{-1}=\frac{1}{m_e^*}+\frac{1}{m_h^*}$ is the
electron and hole effective reduced mass and $E_x^{(l_e^a,l_e^b,l_h^a,l_h^b)}(K_{cm})$ is the exciton binding
energy.  $E_x^{(l_e^a,l_e^b,l_h^a,l_h^b)}(K_{cm})$ is expressible as 
$E_x^{(l_e^a,l_e^b,l_h^a,l_h^b)}(K_{cm})=E-E^{bare}_{gap}-E_e^{(l_e^a,l_e^b)}-E_h^{(l_h^a,l_h^b)}-\frac{\hbar^2K_{cm}^2}{2M}$
where $M=m_e^*+m_h^*$,
$E_e^{(l_e^a,l_e^b)}$ ($E_h^{(l_h^a,l_h^b)}$) is the
confinement energy of the electron (hole) in the
$(l_e^a,l_e^b)$ ($(l_h^a,l_h^b)$) subbands,
$\frac{\hbar^2K_{cm}^2}{2M}$ is the center of
mass motion of the exciton and $V_{1-eff}^{(l_e^a,l_e^b,l_h^a,l_h^b)}(z_r)$ is expressed as     

\begin{equation}
\label{Vexact}
V_{1-eff}^{(l_e^a,l_e^b,l_h^a,l_h^b)}(z_r) =  \int_{-\infty}^{\infty}dx^a_edx_e^bdx^a_hdx^b_h 
|f^{l_e^a}(x_e^a)|^2|f^{l_e^b}(x_e^b)|^2|f^{l_h^a}(x_h^a)|^2|f^{l_h^b}(x_h^b)|^2
V_3(\vec{r}_e,\vec{r}_h).
\end{equation}

\noindent The previous equation is an average of the full 3D
Coulomb potential weighted by the probabilities of finding the
electron and hole along the confined directions.  The
averaging results in
a potential that depends on only the electron and hole coordinates
along the unconfined direction.  In what follows we label the
exitonic states as $X=(x,K_{cm},l_e^a,l_e^b,l_h^a,l_h^b)$.  The label
$x$ is discrete or continuous depending on whether the exciton is
bound or unbound, $K_{cm}$ is the exciton center of mass momentum and
$(l_e^a,l_e^b,l_h^a,l_h^b)$ label the subbands of the electron and
hole that comprise the exciton.
  
To simplify the following calculations, we follow Loudon
\cite{Loud59a} and
model $V_{1-eff}^{(l_e^a,l_e^b,l_h^a,l_h^b)}(z_r)$ as

\begin{equation}
\label{Veff}
V_{1-eff}^{(l_e^a,l_e^b,l_h^a,l_h^b)}(z_r)=V_{1-eff}(z_r) = -\frac{e^2}{\epsilon(|z_r|+z_o)}
\end{equation}

\noindent where $\epsilon$ is the dielectric constant of the material
and $z_o$ is a fit parameter. It is possible, using the confined
electron and hole wavefunctions, to evaluate Eq. \eqref{Vexact}
and find a  value of $z_o$ such that Eq. \eqref{Veff} is a good
approximation to Eq. \eqref{Vexact}.  The inclusion of $z_o$ in Eq. (12) removes the singularity at the origin of the 1D Coulomb
potential and allows us to solve Eq. \eqref{Schro}.  The solution
of Eq. \eqref{Schro} for the 1D excitonic relative motion
wavefunction has been discussed in detail in other works and we refer
the reader to these references\cite{Taka91a,Comb04a}. 

Using the solution to Eq. \eqref{EMAFull}, we evaluate the momentum matrix
element in Eq. \eqref{XR1} between the exciton vacuum $|X_1>=|0>$ and an excited
exciton state $|X_2>=|X>$. We find the exciton-radiation
coupling is expressed as \cite{Chuang}  
\begin{equation}
\label{XR2}
T_{cv}^{X}(K_{cm})=\frac{e}{m_o}\sqrt{\frac{2\pi\hbar}{V_{crystal}\omega n^2
}}\vec{e}\cdot\vec{p}_{cv}\sum_{(l_e^a,l_e^b,l_h^a,l_h^b)}Y_{l_e^a}^{l_h^a}Y_{l_e^b}^{l_h^b}
\phi_x^{*(l_e^a,l_e^b,l_h^a,l_h^b)}(z_r=0)\delta_{K_{cm}=k_e-k_h,q}
\end{equation}

\noindent where 

\begin{equation}
\label{Pmat}
\vec{p}_{cv}=\frac{1}{V_{unit}}\int_{V_{unit}}\,d\vec{r} \,u_c^*(\vec{r})\Biggl(\frac{i}{\hbar}\nabla u_v(\vec{r})\Biggr)
\end{equation}

\noindent is the momentum matrix element of the Bloch functions over a
unit cell of volume $V_{unit}$,

\begin{equation}
\label{SbbndXR}
Y_{l_e^j}^{l_h^j}=\int \,dx^j f^{l_e^j}(x^j)f^{l_h^j}(x^j)
\end{equation}

\noindent is an integral calculating the overlap of the subband
electron and hole wavefunctions across the domain $x^j$ with $j=a,b$,
$\phi_x^{*(l_e^a,l_e^b,l_h^a,l_h^b)}(z_r=0)$ is the envelope function
of the 1D electron-hole relative motion of the type $x$
exciton between subbands $(l_e^a,l_e^b,l_h^a,l_h^b)$ evaluated at zero electron-hole
separation
and $\delta_{K_{cm}=k_e-k_h,q}$ is the Kronecker delta function expressing conservation of
momentum along the unconfined direction.  We point out that
$Y_{l_e^j}^{l_h^j}$ is equal to $\delta_{l_e^j,l_h^j}$ when the electron
and hole experience identical confinement potentials, which we will
use in Section III.

Similarly, we evaluate Eq. \eqref{XL1} between two excited exciton states
 $|X_1>$ and  $|X_2>$. We find the exciton-lattice
 coupling constant is expressible as

\begin{equation}
\label{XL2}
S_{X_1}^{X_2} =
\delta_{K_{cm,1},K_{cm,2}+Q}\Biggl[\delta_{l_{h,2},l_{h,1}}C^{l_{e,2}}_{l_{e,1}}
  B^{X_2}_{X_1}(\alpha_eQ) - \delta_{l_{e,2},l_{e,1}}C^{l_{h,2}}_{l_{h,1}}
  B^{X_2}_{X_1}(\alpha_hQ)\Biggr]
\end{equation}

\noindent where
$\alpha_e=\frac{m_h^*}{M}$, $\alpha_h=-\frac{m_e^*}{M}$, $\delta_{l_{j,2},l_{j,1}}$
implies the $j^{th}$ particle stays in its subbands after the exciton scatters
from the phonon
and the phonon mediated intermediate  excitonic  state
coupling  has been decomposed into a piece related to the exciton
relative motion wavefunction 

\begin{equation}
\label{SbbndXL}
 B^{X_2}_{X_1}(Q) = \int \,dz_r \phi_{x_2}^{*(l_{e,2}^a,l_{e,2}^b,l_{h,2}^a,l_{h,2}^b)}(z_r)e^{-iQz_r}
\phi_{x_1}^{(l_{e,1}^a,l_{e,1}^b,l_{h,1}^a,l_{h,1}^b)}(z_r)
\end{equation}

\noindent and a piece depending on the details of how the possibly
confined phonon \cite{PhonNano}
couples with the confined electron and hole

\begin{equation}
\label{XPhMat}
C^{l_{i,2}}_{l_{i,1}} = \int \, dx_i^a dx_i^b
      f^{*l_{i,2}^a}(x_i^a)f^{*l_{i,2}^b}(x_i^b)
C(x_i^a,x_i^b)
f^{l_{i,1}^a}(x_i^a)f^{l_{i,1}^b}(x_i^b)
\end{equation}

\noindent with $i=e$ or $i=h$. 

We now substitute Eq. \eqref{XR2} and Eq. \eqref{XL2} into Eq. \eqref{Res} and
assume the photon momentum may be neglected when compared to the
electron and hole crystal momentum (the k-selection rule)

\begin{equation}
\label{Wmat}
W_{i \rightarrow f} = \sum_{X_1,X_2}
M_{\substack{(l^a_{e,2},l^b_{e,2},l^a_{h,2},l^b_{h,2}) \\
  (l^a_{e,1},l^b_{e,1},l^a_{h,1},l^b_{h,1})}} ^{cv}
\frac{\phi_{x_2}^{(l^a_{e,2},l^b_{e,2},l^a_{h,2},l^b_{h,2})}(0)
  B^{X_2}_{X_1}(Q=0)  \phi_{x_1}^{*(l^a_{e,1},l^b_{e,1},l^a_{h,1},l^b_{h,1})}(0) }
{(\hbar\omega_l-\hbar\Omega_p-E_b-i\Gamma_b)(\hbar\omega_l-E_{a}-i\Gamma_a)}
\end{equation}

\noindent where the coupling constant is given by 

\begin{equation}
\label{Coupling}
M_{\substack{(l^a_{e,2},l^b_{e,2},l^a_{h,2},l^b_{h,2}) \\
  (l^a_{e,1},l^b_{e,1},l^a_{h,1},l^b_{h,1})}} ^{cv}
 = \frac{e^2(2\pi\hbar)^2 (\vec{e}_s \cdot
  \vec{p}^{*}_{vc})(\vec{e}_l\cdot\vec{p}_{cv})}{m_o^2V_{crystal}\sqrt{\omega_s\omega_l}n_sn_l}
Y_{l_{e,2}^a}^{*l_{h,2}^a}Y_{l_{e,2}^b}^{*l_{h,2}^b}
Y_{l_{e,1}^a}^{l_{h,1}^a}Y_{l_{e,1}^b}^{l_{h,1}^b}
[\delta_{l_{h,2},l_{h,1}}C^{l_{e,2}}_{l_{e,1}}-\delta_{l_{e,2},l_{e,1}} C^{l_{h,2}}_{l_{h,1}}].
\end{equation}

\noindent Finally, substituting Eq. \eqref{Wmat} and Eq. \eqref{Coupling} into
Eq. \eqref{RRSCS} we arrive at 


\begin{equation}
\label{XS}
\frac{d\sigma_{RRS}}{d\Omega} =
\frac{e^4\omega_s^2n_s}{c^4m_o^4\omega_l^2n_l}
\Biggl|\sum_{X_1,X_2}
\bar{M}_{\substack{(l^a_{e,2},l^b_{e,2},l^a_{h,2},l^b_{h,2}) \\
  (l^a_{e,1},l^b_{e,1},l^a_{h,1},l^b_{h,1})}} ^{cv}
\frac{\phi_{x_2}^{(l^a_{e,2},l^b_{e,2},l^a_{h,2},l^b_{h,2})}
  B^{X_2}_{X_1}
  \phi_{x_1}^{*(l^a_{e,1},l^b_{e,1},l^a_{h,1},l^b_{h,1})} }
{(\hbar\omega_l-\hbar\Omega_p-E_b-i\Gamma_b)(\hbar\omega_l-E_{a}-i\Gamma_a)}\Biggr|^2
\end{equation}

\noindent where $\bar{M}$ is defined in a similar manner to
Eq. \eqref{Coupling}, but the constants have been factored out and we have supressed the arguments
in both $\phi_x^l$ and $B_{X_1}^{X_2}$.  The double summation extends
over all intermediate excitonic states.


Equation \eqref{XS} is the central result of the paper and provides
a general expression for calculating the scattering cross-section of
resonant Raman scattering, using third-order time dependent
perturbation theory, from a 1D quantum confined structure
when the intermediate electronic excitations are excitonic in
nature. In Eq. \eqref{XS}, we have factored the numerator into a part
that depends on the relative motion of the 1D excitons
and a part that is a function of both
the electron and hole subband confined wavefunctions and the
Bloch functions from which the exciton is built.  The utility of
this decomposition is that we can focus explicitly on how the
1D exciton influences the Raman scattering
cross-section.  If we are interested in a particular material system,
and wish to obtain an absolute value for its Raman scattering
cross-section, it would be necessary to evaluate all the
system-specific matrix elements in Eq. \eqref{Coupling}.  It should be noted
that Eq. \eqref{XS} allows for the possibility of Raman
scattering between both bound and unbound intermediate excitonic
states. 
 In the next section we calculate the Raman scattering cross-section
for a model 1D system with only a single
conduction and valence subband for which we assume the potentials confining the
electron and hole are identical for simplicity.

 \section{Two-Subband Model}
\begin{figure}[t]
\begin{center}
\includegraphics[width=3.34in]{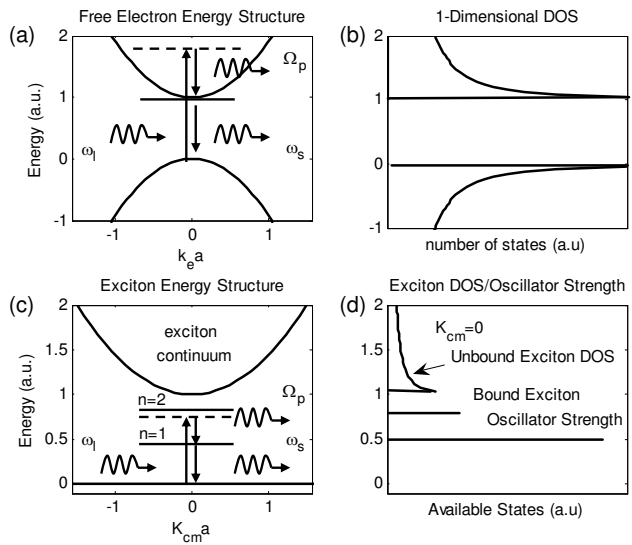}
\end{center}
\caption{The (a) free electron electronic structure, (b) the free
  electron density of states, (c) the excitonic electronic structure and
  (d) the unbound excitonic $K_{cm}=0$ density of states and bound
  exciton oscillator strengths.  In (a), the free electron energies are
  functions of the electron crystal momentum, whereas in (c), the
  excitonic energies are functions of the exciton center of mass
  momentum.  The dispersion associated with the bound excitons
  has not been illustrated since in exciton mediated transtions, only
  $K_{cm}=0$ transitions are allowed due to conservation of momentum.  In addition, in (a) and (c), an outgoing
  Stokes resonance is illustrated where the dashed horizontal line
  corresponds to a virtual electronic state and a solid horizontal
  line corresponds to a real electronic state.  In each case, a photon of energy
  $\hbar\omega_l$ causes an electronic transition to a virtual state,
  followed by the electronic excitation relaxing to a real, electronic
  state by emitting a phonon of energy $\hbar\Omega_p$. Finally, the
  electronic system returns to its ground state by emitting a photon
  of energy $\hbar\omega_s$.  
  This can be compared to the free electron mediated transitions where only
  vertical transitons are allowed, but $k_{e}$ is not constrained to
  be 0.}
\label{Fig2}
\end{figure}

In this section, we give an explicit expression for Eq. \eqref{XS} when
the material system is composed of only a single conduction and
valence subband
($(l_e^a,l_e^b,l_h^a,l_h^b)=(1_e^a,1_e^b,1_h^a,1_h^b)$) and the
electron and hole experience identical confining potentials.  Fig. \ref{Fig2}(a) illustrates the single particle
bandstructure for the system which is assumed to be known.
Using the
single particle bandstructure and wavefunctions, we can solve Eq. \eqref{Schro} for the
exciton energy eigenvalues, and visualize the
electronic structure of our material system in Fig. \ref{Fig2}(c).
Incorporating excitonic effects has resulted in a series of bound
states below the quantum confined system's energy gap of
$E_{gap}^{bare} + E_e^{(1_e^a,1_e^b)}+E_h^{(1_h^a,1_h^b)}$ in addition to
the usual continuum of states above this energy gap.  As discussed in Loudon
\cite{Loud59a}, the internal energy label for bound exciton states is
no longer constrained to a set of postive integers, but is describable
by a  set of positive real numbers.  Within
this two-subband approximation, we can simplify Eq. \eqref{XS} as follows

\begin{equation}
\label{XS2band}
\frac{d\sigma_{RRS}}{d\Omega}(\omega_l;\omega_s=\omega_l-\Omega_p) =
R|L_{bound}(\omega_l;\omega_s=\omega_l-\Omega_p)+L_{unbound}(\omega_l;\omega_s=\omega_l-\Omega_p)|^2
\end{equation}  

\noindent where $L_{bound}$ and $L_{unbound}$ are functions that
characterize the influence that the
bound and unbound excitons have on the Raman scattering cross-section and
we have collected all the constants of Eq. \eqref{XS} in $R$.  In this two-subband model, transitions between intermediate exciton states, with
different values of $x$, are forbidden by $B_{X_1}^{X_2}$ in
Eq. \eqref{SbbndXL} when we assume the k-selection rule and neglect the phonon momentum $Q$.
For transitions between different excitonic states to be possible, there must be 
non-zero overlap between the two intermediate
states that participate in the Raman scattering process.  Overlap can arise if the excitons are derived from subbands
with different effective masses and therefore have
unequal Bohr radii.
  
Specific expressions for $L_{bound}$ and $L_{unbound}$ are

\begin{equation}
\label{Bnd}
L_{bound}(\omega_l;\omega_s=\omega_l-\Omega_p) = \sum_{n}\frac{|\phi_{n}(0)|^2}
{(\hbar\omega_s-E_{gap}+\frac{R^*}{n^2}-i\Gamma_b)
(\hbar\omega_l-E_{gap}+\frac{R^*}{n^2}-i\Gamma_a)} 
\end{equation}

\noindent and

\begin{equation}
\label{UBnd}
L_{unbound}(\omega_l;\omega_s=\omega_l-\Omega_p) = \int_{E_{gap}}^{E_c}
  \,dE\frac{|\phi_{E}(0)|^2}
{\sqrt{E_{gap}-E}(\hbar\omega_s-E_{gap}-E-i\Gamma_b)
(\hbar\omega_l-E_{gap}-E-i\Gamma_a)}
\end{equation}

\noindent where $R^*$ is the effective exciton Rydberg, $E_c$ is an
energy cutoff to the above integral and the summation in Eq. \eqref{Bnd}
includes only even envelope functions.  In Eq. \eqref{Bnd}
(Eq. \eqref{UBnd}), the sum over all intermediate states has been reduced to
a sum (an integral)
over the label associated with the internal energy of the exciton.

With Eqs. \eqref{Bnd} and \eqref{UBnd}, we use both the bound and
unbound exciton relative motion wavefunctions to evaluate the
lineshape of the exciton mediated Raman scattering
cross-section.     
In evaluating the bound and unbound wavefunctions with Eq. \eqref{Schro}, we have freedom in how we choose $z_o$,
the parameter introduced to make the 1D Coulomb potential
finite at the origin.  The closer $z_o$ is to zero, the greater the
1D character of the problem we are solving.  Since the majority of
physical systems exhibiting properties characteristic of a
1D system typically are not truly 1D, finite values of
$z_o$ are physically reasonable.  For example, in a SWNT, the tube
radius provides a natural value for $z_o$.  The procedure to correctly
determine $z_o$ is to select $z_o$ such that Eq. \eqref{Veff}
approximates Eq. \eqref{Vexact}. 

\begin{figure}[t]
\begin{center}
\includegraphics[width=3.34in]{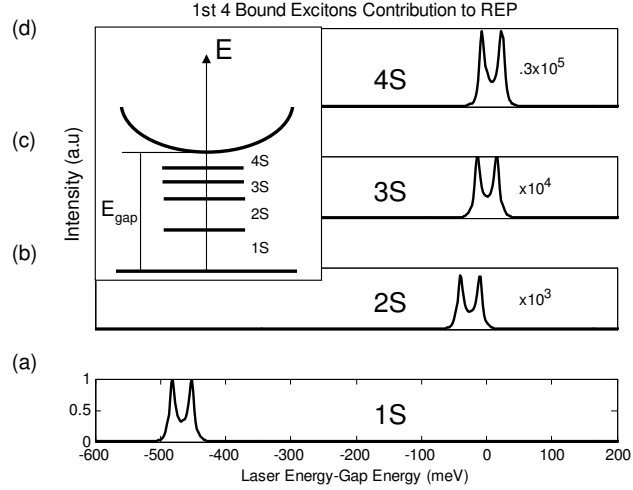}
\end{center}
\caption{The (a) 1S,(b) 2S,(c) 3S and (d)
  4S bound exciton REP.  In all figures,
  $z_o/a_X=.2$, $\Gamma=5$ meV, $R^*=100$ meV and $\hbar\Omega_p=32$ meV.  The double
  peak structure in the REP is due to both an incoming resonance, when
  $\hbar\omega_{l}=E_{gap}$, and an outgoing resonance, when $\hbar\omega_s=E_{gap}$.} 
\label{Fig3}
\end{figure}

Before evaluating the full lineshape function Eq. \eqref{XS2band}, we will focus on the
individual contributions of the bound and unbound excitons.  Fig. \ref{Fig3}
illustrates the contributions of the first four bound excitons to the bound lineshape
function Eq. \eqref{Bnd} as a function of laser energy $\hbar\omega_l$.
The two peak structure apparent in the 1phRRS cross-section mediated by
each bound exciton is a result of an incoming and an outgoing resonance.
These resonances are associated with the incoming (laser) or outgoing
(scattered) photon being coincident in energy with the bound exciton
internal energy. 
The dependence of the 1phRRS on excitation frequency is commonly
referred to as a Resonance Excitation Profile (REP) and we will adhere
to this terminology.
In Fig. \ref{Fig3}, all exciton REPs are normalized relative to the
ground state exciton mediated transition (the top figure).
To
calculate each REP, we assumed a dephasing of
$\Gamma_a=\Gamma_b=5$ meV, a phonon of energy $\hbar\Omega_p=32$ meV and
we set $E_{gap}=E^{bare}_{gap}+E_e^{(1_e^a,1_e^b)}+E_h^{(1_h^a,1_h^b)}$ to
be the zero of the energy scale.  In addition, we chose the Rydberg
$R^*=100$ meV.
 The reason for choosing this value is that it sets the ground
state exciton binding energy to approximately 500 meV, a value determined in
recent measurements on SWNTs \cite{Hein05a}.   It is clear in Fig.
\ref{Fig3} that the ground state exciton dominates the bound exciton
mediated lineshape function.   The contribution to the REP
from the ground state exciton is nearly \textit{3 orders of magnitude}
stronger than that of the next bound excited exciton.

\begin{figure}[t]
\begin{center}
\includegraphics[width=3.34in]{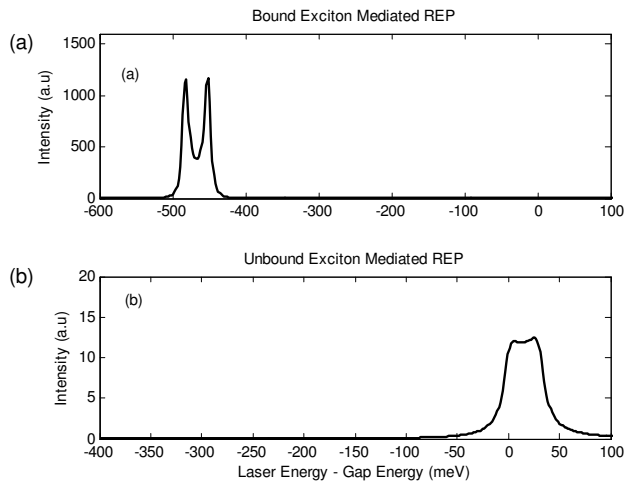}
\end{center}
\caption{Comparison of the relative strength of the (a) bound exciton mediated
 REP with the (b) unbound exciton mediated REP. In (a) and (b),
  $z_o/a_X=0.2$, $\Gamma=5$ meV, $R^*=100$ meV and $\hbar\Omega_p=32$ meV.}
\label{Fig4}
\end{figure}

Next, with the same set of parameters, we compare, in absolute terms, the unbound exciton
contribution to the exciton mediated REP with the contribution of the strongest bound
exciton.  To make this comparison we plot Eq. \eqref{Bnd} in Fig. \ref{Fig4}(a) and
Eq. \eqref{UBnd} in Fig. \ref{Fig4}(b). 
The strength of the bound exciton mediated
REP is nearly \textit{ 2 orders of magnitude} larger than the unbound exciton.
 It is clear that
$|\phi_{x}(0)|^2$ strongly influences how efficiently a given exciton
  can mediate the 1phRRS process. Similar to it effect in 1D absorption
  \cite{Taka91a}, $|\phi_{x}(0)|^2$ acts to suppress the contribution of the unbound
  exciton to the exciton mediated REP.  

With an understanding of how both the bound and unbound exciton
contribute individually to the REP, we now evaluate Eq. \eqref{XS2band}
for the full exciton mediated 1phRRS REP.  Since we will compare our result with the REP of
the free electron-hole (FEH) theory for 1phRRS, we quote the
result for the free electron-hole theory Raman scattering cross-section \cite{CardScatt}
\begin{equation}
\label{FEHXS}
\frac{d\sigma^{FEH}_{RRS}}{d\Omega}(\omega_l;\omega_s=\omega_l-\Omega_p)=
\frac{e^4\omega_s^2n_s}{c^4m_o^4\omega_l^2n_l}
\frac{\pi^2}{\hbar^2\Omega^2_p}
\Biggl|\frac{1}{\sqrt{(\hbar\omega_l-\hbar\Omega_p-E_{gap}}}-\frac{1}{\sqrt{i\Gamma_b)(\hbar\omega_l-E_{gap}-i\Gamma_a)}}\Biggr|^2.
\end{equation}

\begin{figure}[t]
\begin{center}
\includegraphics[width=3.34in]{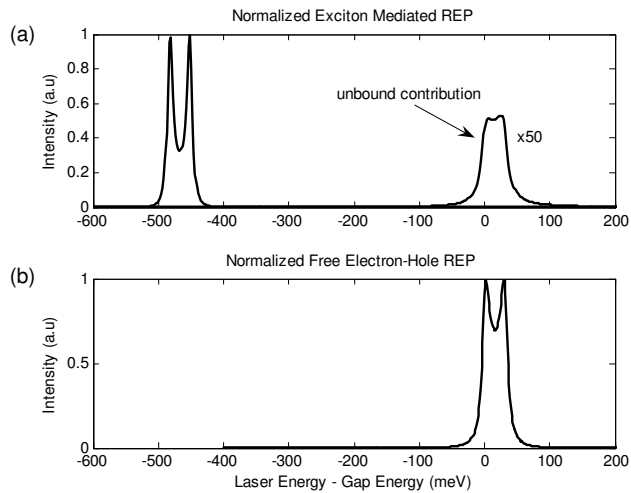}
\end{center}
\caption{Comparison of the normalized (a)  exciton mediated
 REP with the (b) Free electron-hole mediated REP. In (a) and (b),
  $\Gamma=5$ meV, $R^*=100$ meV and $\hbar\Omega_p=32$ meV. In addition,
 in (a), $z_o/a_X=0.2$. Notice in (a) the small contribution of
 the unbound excitons.}
\label{Fig5}
\end{figure}

\noindent Figure \ref{Fig5} compares  Eq. \eqref{XS2band} with
  Eq. \eqref{FEHXS}.
It is
immediately clear from Fig. \ref{Fig5}(a) that the energies at which the system strongly Raman scatters
incident laser light is shifted to the lowest bound exciton.  Nonetheless,
there is a small feature at the band gap associated with light
Raman scattered using unbound exciton intermediate states.  In
addition, the incoming and outgoing peaks are more pronounced in the
exciton mediated REP as compared to the FEH REP.  Practically, though, such
small qualitative differences in the REP are likely to be experimentally undetectable.  Besides the gross
shift in energy, the REPs generated by both theories appear
quite similar.  

\begin{figure}[t]
\begin{center}
\includegraphics[width=3.34in]{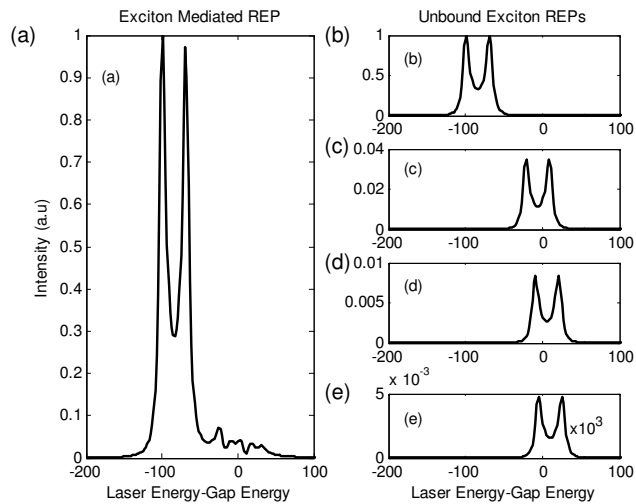}
\end{center}
\caption{The (a) exciton mediated REP and in (b)-(e), the REPs
of the first four bound excitons. In (a)-(e),
  $z_o/a_X=1$, $\Gamma=5$ meV, $R^*=100$ meV and $\hbar\Omega_p=32$ meV.  Notice the
  structure present at the location of the free electron bandgap. }
\label{Fig6}
\end{figure}

At this point it is important to recall that, in solving
Eq. \eqref{Schro}, we set the ratio of the potential cutoff to
the exciton Bohr radius, ${z_o}/{a_X}$,  equal to 0.2.  If we further
reduce $z_o \rightarrow 0$, the lowest bound exciton will only become more
dominant in mediating the REP.  With this in mind,
  we finally
investigate the dependence of the exciton mediated REP on $z_o/a_X$.
In particular, we now set $z_o/a_X=1$.  Physically, the Bohr radius of
the exciton is equal to the cutoff of the 1D Coulomb
potential and,  as $z_o$ increases, the system becomes less
1D.  In Fig. \ref{Fig6}, we keep all input parameters from Fig. \ref{Fig5} fixed except
$z_o/a_X$ is changed to 1.

First, the higher excited bound excitons (Fig. \ref{Fig6}(b)-\ref{Fig6}(e)) make larger individual
contributions to the exciton mediated REPs.  When we examine the full
exciton mediated REP, we find there is some structure in the vicinity
of the the free electron band gap. We can understand this structure as
follows.  As the binding energy of the
ground state exciton decreases, the relative strength of the
scattering process mediated
by the lowest bound exciton (as compared to the other bound excitons)
also decreases.  In addition, 
the REPs that we have attributed to
the various bound excitons begin to overlap.  In fact, as the overlap 
increases, the REP at a fixed $\hbar\omega_l$ is the result of a quantum
interference between all excitonic pathways that can contribute
effectively to the scattering process.

\section{Conclusion}  

We have developed a general theory for calculating the exciton mediated one phonon
resonant Raman scattering cross-section for 1D quantum
confined systems (Eq. \eqref{XS}).  In studying a model two-subband system, we
found the exciton was strongly bound when $z_o$ is small
compared to the exciton Bohr radius $a_X$.  In this limit
of small $z_o$, the ground state
exciton dominates the 1phRRS REP.   The contribution to the REP from
unbound excitons with energies in the range of the single particle
gap energy is quenched.  The quenching is similar in origin to the
suppression of the exciton mediated absorption coefficient at
energies above a 1D material system's bare energy
gap; the ground state exciton carries all the spectral weight in the
transition \cite{Taka91a}.      
As the Coulomb potential cutoff $z_o$ is increased, the ground
 state becomes more weakly bound, and we found
 that both the higher excited bound and unbound excitons begins to contribute to the 1phRRS
REP.  As $z_o$ approaches the excitonic Bohr radius $a_X$, the REP, at a fixed
laser frequency, is the result of a quantum
interference between all contributing intermediate excitonic pathways.
 The interferences lead to a complicated structure at the single
 particle energy gap.  The shape of the 1phRRS REP provides a qualitative
 indicator of how much spectral weight the ground state exciton
 carries in mediating the 1phRRS process.   

The sensitivity of the 1phRRS REP to changes in
 the potential cutoff is a physically important effect.  Although the
 cutoff is introduced to make tractable the analytic solution of the
 Schr\"{o}dinger equation with the 1D Coulomb potential
 tractable, the majority of physically realizable quantum confined
 1D systems are more accurately described as quasi-1D.  For example, both semiconductor quantum wires and
 SWNTs have finite spatial extent in the directions perpendicular to
 the direction of unconfined motion.  The finite extent in these
 spatial directions lends
 itself naturally to the introduction of a cutoff in the Coulomb potential.  

In applying our results to specific material systems, such as
semiconductor quantum wires or SWNTs, it is necessary to
evaluate matrix elements specific to each material system. Though not
discussed in this work,
we also observe that if we allow
for the possibility of more than two subbands, it becomes possible to
observe true double resonances in the 1phRRS REP.  Specifically, the
phonon could scatter the intermediate exciton between two real, bound
states.  We leave such investigations to future work.

\begin{acknowledgments}
This work was supported by Air Force Office of Scientific Research
under Grant No. MURI F-49620-03-1-0379, by NSF under Grant No. NIRT
ECS-0210752 and a Boston University SPRInG grant.  The authors would like to thank Ernie Behringer for
reviewing the manuscript.  
\end{acknowledgments}


\end{document}